\documentclass{ws-procs9x6}
\usepackage{graphicx}
\usepackage{amsmath}

\begin{document}

\title{Controlled photon emission and Raman transition experiments with a single trapped atom. }

\author{M.~P.~A. JONES, B. DARQUI\'{E}, J. BEUGNON, J. DINGJAN, S. BERGAMINI, Y. SORTAIS, G. MESSIN, A. BROWAEYS and P. GRANGIER}

\address{Laboratoire Charles Fabry de l'Institut d'Optique \\
B\^{a}t. 503 Centre Universitaire \\ 
91403 Orsay, France\\ 
E-mail: Matt.Jones@iota.u-psud.fr}

\maketitle

\abstracts{We present recent results on the coherent control of an optical transition in a single rubidium atom, trapped
in an optical tweezer. We excite the atom using resonant light pulses that are short (4\,ns) compared with the lifetime
of the excited state (26\,ns). By varying the intensity of the laser pulses, we can observe an adjustable number of Rabi
oscillations, followed by free decay once the light is switched off. To generate the 
pulses we have developed a novel laser system based on frequency doubling a telecoms laser diode at 1560\,nm. By setting
the laser intensity to  make a $\pi$ pulse, we use this coherent control to make a high quality triggered source of
single photons.  We obtain an average single photon rate of $9600\,\mathrm{s}^{-1}$ at the detector. Measurements of
the second-order temporal correlation function show almost perfect antibunching at zero delay. In addition,
we present preliminary results on the use of Raman transitions to couple the two hyperfine levels of the ground state of
our trapped atom. This will allow us to prepare and control a qubit formed by two hyperfine sub-levels. }

\section{Introduction}

In order to use a particular physical system for quantum computation, it is necessary to be able to perform single-qubit
operations such as rotations, and two-qubit operations such as  controlled-not gates. These two basic building blocks
can then be concatenated to realise any other desired logical operation.

Neutral atoms have been proposed as a candidate physical system for quantum information processing. For example, the
alkali atoms have two hyperfine levels in the ground state which can be used to make a qubit with very long coherence
times. Single-qubit operations can be realised by using microwaves to drive the hyperfine transition directly, or by
using a two-photon Raman transition. Recently, addressable single-qubit operations have been successfully demonstrated on
a ``quantum register'' of trapped atoms using microwaves\cite{meschede}.

So far, individually addressed two-qubit gates have not been demonstrated with neutral atoms. 
Deterministic gates generally require a strong interaction between the particles that are used to carry the physical
qubits\cite{Zoller},
such as the Coulomb interaction between trapped ions. Promising results have been obtained on entangling neutral atoms
using cold controlled collisions in an optical lattice\cite{bloch}, but the single-qubit operations are difficult to perform in such a system.

 Another approach is to bypass the requirement for a direct
interaction between the qubits, and use instead an interference
effect and a measurement-induced state projection to create the
desired operation\cite{KLM,Dowling}. An interesting recent
development of this idea is to use photon detection events for
creating entangled states of two atoms\cite{protsenko02b,simon,duan03,barrett05}. This provides ``conditional"
quantum gates, where the success of the logical operation is
heralded by appropriate detection events. Such schemes can be
extended to realize a full controlled-not gate, or more generally to implement conditional unitary
operations\cite{protsenko02b,Beige}. These proposals require the controlled emission of indistinguishable photons by
the two atoms.

In this paper we describe our recent experiments\cite{our_science} on triggered single-photon emission from a single
rubidium atom trapped at
the focal point of a high-numerical-aperture lens (N.A. = $0.7$). We show that we have full control of the optical
transition by observing Rabi oscillations. Secondly, we describe preliminary results on the use of Raman transitions to
couple the two ground state hyperfine levels of the trapped atom with a view to performing single-qubit operations.

\section{Trapping single atoms}

\begin{figure}
\centering
\includegraphics{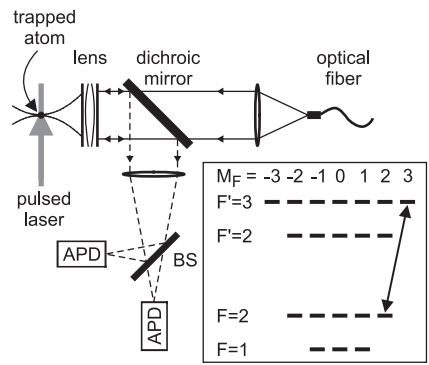}
\caption{Schematic of the experiment. The same lens
is used to focus the dipole trap and collect the fluorescence
light. The fluorescence is separated by a dichroic mirror and
imaged onto two photon counting avalanche photodiodes (APD),
placed after a beam-splitter (BS). The insert shows the relevant
hyperfine levels and Zeeman sub-levels of rubidium $87$. The cycling
transition is shown by the arrow. \label{setup}}
\end{figure}

We trap the single rubidium 87 atom at the focus of the lens using a far-detuned optical dipole trap (810 nm), loaded
from an optical molasses. The same lens is also used to collect the fluorescence emitted by the atom (Fig. \ref{setup}),
which is then detected using a photon counting avalanche photodiode.  The overall detection and collection efficiency
for the light emitted from the atom is measured to be $0.60 \pm 0.04\%$. A crucial feature of our experiment is the
existence of a ``collisional blockade" mechanism\cite{Schlosser01,Schlosser02} which allows only one atom at a time to be stored in
the trap: if a second atom enters the trap, both are immediately ejected. In this regime the atom statistics are
sub-Poissonian and the trap contains either one or zero (and never two) atoms, with an average atom number of $0.5$. The
experimental apparatus is described in more detail in references\cite{Schlosser01,Schlosser02}.

\section{Triggered single photon emission}

We excite the atom with $4$\,ns pulses of laser light, resonant with the \mbox{$S_{1/2},\, F=2 \rightarrow P_{3/2},\,
F^{\prime} = 3$} transition at $780.2$\,nm (see insert in figure \ref{setup}). These pulses are much shorter than the
$26$\,ns lifetime of the upper state.
The pulsed laser beam is $\sigma^+$-polarized with respect to  the quantisation axis which is defined by a $4.2$\,G
magnetic field. The trapped atom is optically pumped into the $F=2,\, m_F =+2$ ground
state by the first few laser pulses. It then cycles on the \mbox{$F=2,\, m_F=+2 \rightarrow F^{\prime}=3,\, m_F^{\prime}=+3$}
transition, which forms a closed two-level system emitting $\sigma^+$-polarized photons. Impurities in the polarisation
of the pulsed laser beam with respect to the quantisation axis, together with the large bandwidth of the exciting pulse
($250$ MHz), result in off-resonant excitation to the $F'=2$ upper state, leading to possible de-excitation to the $F=1$
ground state. To counteract this, we add a repumping laser resonant with the $F=1
\rightarrow F' = 2$ transition. We have checked that our two-level description is still valid in the presence of the
repumper by analyzing the polarisation of the emitted single photons\cite{our_science}. 

\begin{figure}
\centering
\includegraphics[height=5cm]{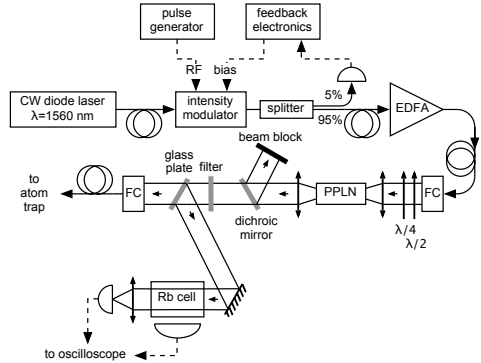}
\caption{Schematic of the pulsed laser system. FC: Fibre Coupler\label{laser_system_diagram}}
\end{figure}

To generate these short laser pulses, we have developed a novel laser system\cite{tech_paper}, shown in figure \ref{laser_system_diagram}, based on frequency
doubling pulses at the
1560 nm telecoms wavelength. The pulses are generated using an electro-optic
modulator to chop
the output of a continuous-wave diode laser. A commercial fibre amplifier is used to boost the peak power of the pulses
prior to the doubling stage. The laser, modulator, and amplifier are all standard telecommunications components.
The frequency doubling is performed in a single pass using a heated periodically-poled lithium niobate (PPLN) crystal. We
monitor the centre frequency using the fluorescence in a rubidium vapour cell, and tune the system using the temperature of the diode laser. The repetition rate of the
source is $5$ MHz, and we obtain peak powers of up to 12\,W. 

\subsection*{Rabi oscillations}

For a two-level atom and exactly resonant square light pulses of fixed
duration $T$, the probability for an atom in the ground state to
be transferred to the excited state is $\sin^2(\Omega T/2)$, the
Rabi frequency $\Omega$ being proportional to the square root of
the power. Therefore the excited state population and hence the
fluorescence rate oscillates as the intensity is increased. To
observe these Rabi oscillations, we illuminate the trapped atom
with the laser pulses during $1$ ms. We keep the length of each laser
pulse fixed at $4$\,ns, with a repetition rate of $5$ MHz, and
measure the total fluorescence rate as a function of the laser
power. The Rabi oscillations are clearly visible on our results
(see Fig. \ref{inter}a). From the height of the first peak and the calibrated
detection efficiency, we derive a maximum
excitation efficiency per pulse of $95\pm 5\%$.

The reduction in the contrast of the oscillations at high laser
power is mostly due to fluctuations of the pulsed laser peak
power. This is shown by the theoretical curve in Fig. \ref{inter}a,
based on a simple two-level model. This model shows
that the $10\%$ relative intensity fluctuations that we measured on
the laser beam  are enough to smear out the oscillations as
observed.

The behaviour of the atom in the time domain can be studied by
using time resolved photon counting techniques to record the
arrival times of the detected photons following the excitation
pulses, thus constructing a time spectrum. By adjusting the laser
pulse intensity, we observe an adjustable number of Rabi
oscillations during the duration of the pulse, followed by the
free decay of the atom once the laser has been turned off. The
effect of pulses close to $\pi$,  $2\pi$ and $3\pi$ are displayed
as inserts on Fig. \ref{inter}a. 

\subsection*{Single photon emission}

In order to  use this system as a single photon source, the laser
power is set to  realize a $\pi$ pulse. To maximise the number of
single photons emitted before the atom is heated out of the trap,
we use the following sequence. First, the presence of an atom in
the dipole trap is detected in real-time using its fluorescence
from the molasses light. Then, the magnetic field is switched on
and we trigger an experimental sequence that alternates
$115\,\mu$s periods of pulsed excitation with $885\,\mu$s periods
of cooling by the molasses light. After $100$ excitation/cooling cycles,
the magnetic field is switched off and the  molasses is turned
back on, until a new atom is recaptured and the process begins
again. On average, three atoms are captured per second under these
conditions. The average count rate during the excitation is
$9600$~s$^{-1}$, with a peak rate of $29000$~s$^{-1}$.

\begin{figure}[ht]
\centering
\includegraphics[width=11cm]{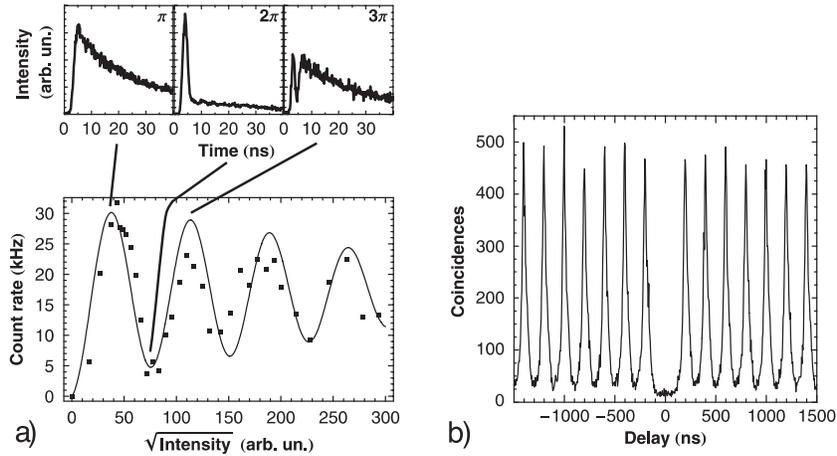}
\caption{a) Total count rate as a function of average power for a pulse length of 4\,ns. Rabi oscillations are clearly
visible. The inset shows the time resolved fluorescence signal for $\pi$, $2\pi$ and $3\pi$ pulses. b) Histogram of the
time delays in the HBT experiment. The peak at zero delay is absent, showing that the source is emitting single photons.
\label{inter}}
\end{figure}

To characterize the statistics of the emitted light, we measure
the second order temporal correlation function, using a Hanbury
Brown and Twiss (HBT) type set-up. This is done using the beam splitter
in the imaging system (Fig. \ref{setup}), which sends the fluorescence light
to two photon-counting avalanche photodiodes that are connected to
a high-resolution counting card in a
start-stop configuration (resolution of about 1 ns). The card is
gated so that only photons scattered during the periods
of pulsed excitation are counted, and the number of coincidence
events is measured as a function of delay. The histogram obtained
after $4$ hours of continuous operation is displayed in Fig. \ref{inter}b, and
shows a series of spikes separated by the period of the excitation
pulses ($200$ ns). The $1/e$ half width of the peaks is \mbox{$27 \pm 3$\,
ns}, in agreement with the lifetime of the upper state. 

After correction for a small flat background due to stray laser light and dark counts, the integrated residual area
around zero delay is \mbox{$3.4\%\, \pm\, 1.2\%$} of the area of the other peaks. This is due to a small
probability for the atom to emit a photon {\it during} the pulse, and be re-excited and emit a second photon. From our calculation\cite{our_science}, the probability for the atom to emit two photons per pulse is 0.018. 

\section{Raman transitions - towards a qubit}

To drive Raman transitions between the two hyperfine levels, two phase-coherent laser beams with a frequency difference
equal to the hyperfine transition frequency are required. In our experiment we use the dipole trap itself as one of the
Raman beams. The large detuning minimises problems due to spontaneous emission during the Raman pulse. Due to the very
high intensity at the waist of the dipole trap, high two-photon Rabi frequencies can still be obtained. The second
Raman beam is generated using two additional diode lasers that are phase-locked to the dipole trap by injection
locking. The 6.8 \,GHz frequency offset is generated by modulating the current of one of the diode lasers at 3.4\,GHz. 

To drive Raman transitions for qubit rotations, we superimpose the second Raman beam with the dipole trap beam. As the
beams are co-propagating, this makes the transition insensitive to the external state of the atom. The beam is linearly
polarised orthogonal to the dipole trap. With the quantisation axis defined by a 4.2\,G magnetic field as described
above, this means that we can drive $\pi /\sigma^{+}$ or $\pi /\sigma^{-}$ transitions.

\begin{figure}[t]
\centering
\includegraphics{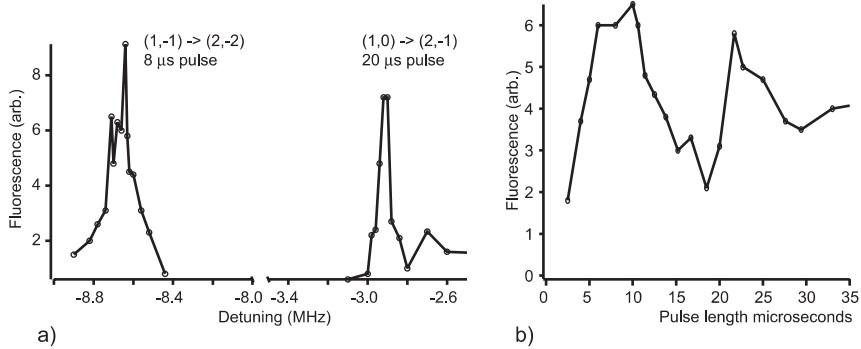}
\caption{a) Fluorescence as a function of detuning between the Raman beams with an applied magnetic field of 4.2\,G. 
Zero
detuning corresponds to the hyperfine splitting in zero applied field. Two of the possible transitions between magnetic
sublevels are shown: \mbox{$F=1,m_{F}=-1 \rightarrow F=2,m_{F}=-2$ (left) and $F=1,m_{F}=0 \rightarrow F=2,m_{F}=-1$} (right). b)
Fluorescence as a function of the duration of the Raman pulse.
\label{raman_peaks}}
\end{figure}

We perform spectroscopy of the transitions as follows. The atom is prepared in the $F=1$ hyperfine level by switching
off the repump light 1\,ms before the molasses light. This process populates all of the magnetic sublevels. Then we
transfer the atom to the $F=2$ hyperfine level by pulsing on the second Raman beam. The population in the $F=2$ level is
detected using the fluorescence from a resonant probe beam. By measuring the fluorescence as a function of the frequency
difference between the two Raman beams we obtain spectra such as those shown in figure \ref{raman_peaks}a. The width of these
peaks is limited
by the length of the Raman pulse in each case. 

We have also made a preliminary observation of Rabi oscillations. The detuning between the Raman lasers was set resonant
with the \mbox{$F=1,m_F=-1 \rightarrow F=2,m_F=-2$} transition, and we measured the population in $F=2$ as a function of the
pulse length. The results are shown in figure \ref{raman_peaks}b. The measured Rabi frequency is $\Omega = 2 \pi \times 65$\,kHz, with
a power of only 60\,nW in the second Raman beam. As we have 10\,mW available, we should be able to attain Rabi
frequencies in the MHz range.

\section{Conclusions and outlook}
 We have realized a high quality source of single photons based on the coherent excitation of an optically trapped single atom. Preliminary 
results on using Raman transitions to couple the two hyperfine ground states of
the trapped atom have also been obtained. In the future, we would like to extend these experiments to several atoms. In previous work we
have shown that we can create two-dimensional arrays of dipole traps separated by distances of several microns, each
containing a single atom\cite{slmpaper}.  Our goal is to see whether we can realise single-qubit and two-qubit operations in such arrays.

\section*{Acknowledgments}
This work
was supported by the European Union through the IST/FET/QIPC project ÒQGATESÓ and
the Research Training Network ÒCONQUESTÓ. M. Jones was supported by EU Marie Curie fellowship MEIF-CT-2004-009819

\end{document}